\documentclass[onecolumn]{aastex6}

\begin{document}

\title{A Unified Model for Repeating and Non-repeating Fast Radio Bursts}

\author{Manjari Bagchi\altaffilmark{1}}
\affil{The Institute of Mathematical Sciences (IMSc-HBNI),  \\
4th Cross Road, CIT Campus, Taramani,  \\
Chennai 600113, India}

\altaffiltext{1}{manjari@imsc.res.in}

\begin{abstract}

The model that fast radio bursts are caused by plunges of asteroids onto neutron stars can explain both repetitive and non-repetitive  bursts. If a neutron star passes through an asteroid belt around another star, there would be a series of bursts caused by series of asteroid impacts. Moreover, the neutron star would cross the same belt repetitively if it is in a binary with the star hosting the asteroid belt, leading to a repeated series of bursts. I explore the properties of neutron star binaries which could lead to the only known repetitive fast radio burst so far (FRB121102). In this model, the next two epochs of bursts are expected around 27-February-2017 and 18-December-2017.  

On the other hand, if the asteroid belt is located around the neutron star itself, then a chance fall of an asteroid from that belt onto the neutron star would lead to a non-repetitive burst. Even a neutron star grazing an asteroid belt can lead to a non-repetitive burst caused by just one asteroid plunge during the grazing. This is possible even when the neutron star is in a binary with the asteroid hosting star, if the belt and the neutron star orbit are non-coplanar.

\end{abstract}


\section{Introduction} \label{sec:intro}

Since the first discovery a decade ago by \citet{lorimer07}, Fast Radio Bursts (FRBs) gave rise to a plethora of models. Some of those are of catastrophic in origin, like mergers of two neutron stars \citep{tot13,wyw16}, mergers of two white-dwarfs \citep{kim13}, mergers of two black holes when at least one is charged \citep{zhang16}, collapses of supra-massive neutron stars into black holes \citep{fr14}, etc., while some others are non-catastrophic like the magnetospheric activity of neutron stars \citep{pp13, pl15, katz16}, collisions between neutron stars and asteroids \citep{gh15, dwwh16}, flares from stars in the Galaxy \citep{lsm14}, etc.

The repetitive nature of FRB121102 \citep{Spsh16, Scsh16} has ruled out the catastrophic models, unless this event is of an origin distinct from all other FRBs. Localization of this event to within a dwarf galaxy at a redshift of $z = 0.19273(8)$ has excluded models involving Galactic origin as well \citep{clw17, mph17, tbc17}.

The fall of an asteroid of a mass of a few $10^{18}$ gm can cause an FRB \citep[and references therein]{dwwh16}. When an asteroid comes close to a magnetized neutron star, a very large electric field is induced parallel to the magnetic field of the neutron star. This induced electric field detaches electrons from the surface of the deformed asteroid and accelerates these electrons to ultra-relativistic energies. Curvature radiation from these ultra-relativistic electrons moving along the magnetic field lines produces FRBs. This model has the potential to explain a diverse population of FRBs depending on the nature of the impact. If there is an asteroid belt around a neutron star, then a chance fall of an asteroid from that belt onto the neutron star would lead to a single burst. \citet{gh15} argued that one neutron star is likely to face only one such asteroid impact, i.e. a non-repeating burst from a particular source. Later, \citet{dwwh16} demonstrated that a neutron star traveling through an asteroid belt would face several asteroid impacts resulting in a series of bursts. Although \citet{dwwh16} mainly concentrated on the phenomenon of isolated neutron stars moving in a galaxy and passing through asteroid belts around other stars, they also mentioned that if the neutron star comes too close to the asteroid-hosting star, then it might get captured by the other star, forming a binary. In this case, the neutron star would cross the asteroid belt repetitively leading to repeating series of bursts and this might be the case for FRB121102. In the present letter, I extend this model. Note that, it is not essential that the neutron star binary would form only via the capture process, it is also possible that the asteroid belt was created during the evolution of the binary. The neutron star would cross (twice in each orbital revolution) the asteroid belt around its stellar companion if the radius of the asteroid belt is smaller than the apastron distance of the neutron star, as well as the orbit of the neutron star is eccentric (at least mildly), as no crossing is possible in case of two concentric circular orbits. When both of the above conditions are satisfied, there would be more than one plunges of asteroids onto the neutron star giving rise to a series of FRBs when the neutron star is inside the asteroid belt followed by a quiescent period (due to the absence of such plunges) when the neutron star is out of the belt; and the whole process would repeat due to the orbital motion of the neutron star. I elaborate this special situation in section \ref{sec:model}.

Sections \ref{sec:applicationmodel} and \ref{sec:res} demonstrate the application of the model for the case of FRB121102 including possible system parameters which would agree with observations. Finally, in section \ref{sec:discuss}, I generalize the model and discuss how non-repetitive FRBs can also be explained within this model. 

\section{Asteroid infalls onto a neutron star passing through an asteroid belt} \label{sec:model}

The rate of infalls of asteroids onto a neutron star passing through an asteroid belt, i.e. the rate of FRBs can be written as \citep{dwwh16}: $\mathcal R \sim 1.25 \times 10^{10} \, R_{\rm ns, 6} \, (M_{\rm ns}/{1.4 \rm M_{\odot}}) \, v_{\rm ns, 7}^{-1} \, N_{a} \, (\eta_1/0.2)^{-1} \, (\eta_2/0.2)^{-1} \, (R_{\rm belt}/2{\rm AU})^{-3} ~{\rm h}^{-1}$ where $R_{\rm ns, 6}$ is the radius of the neutron star in the unit of $10^6$ cm, $v_{\rm ns, 7}$ is the speed of the neutron star in the unit of $10^{7} ~ {\rm cm \, s^{-1}} $, $N_{a}$ is the total number of asteroids in the belt, $\eta_1 R_{\rm belt}$ and $\eta_2 R_{\rm belt}$ are the thickness and width of the belt, $R_{\rm belt}$ is the inner radius of the belt and $\eta_1$ and $\eta_2$ are two fractional numbers. $R_{\rm ns, 6}=1$ is the standard value for the radius of neutron stars.  \citet{dwwh16} used $v_{ns, 7}=2$ as the speed of isolated neutron stars moving in their host galaxies. One needs to replace this value by the orbital speed of the neutron star in order to estimate the rate of asteroid infalls during the passage of the neutron star through an asteroid belt around its binary stellar companion. The orbital speed of the neutron star can be written as:

\begin{equation}
v_{\rm b, ns} (f) = \sqrt{\frac{G (M_{\rm ns} + M_{\rm com})}{a_{\rm ns} (1-e^2)}} \left[1+2 e \cos f +e^2  \right]^{1/2}.
\label{eq:vns}
\end{equation} where $G$ is the gravitational constant, $f$ is the true anomaly of the neutron star, $a_{\rm ns}$ and $e$ are the semi-major axis and the eccentricity of the orbit of the neutron star respectively. $M_{\rm ns}$ and $M_{\rm com}$ are the masses of the neutron star and the companion. $a_{\rm ns}$ is related $P_{\rm b}$ as $a_{\rm ns} =\frac{M_{\rm com}}{(M_{\rm ns} + M_{\rm com})}  \left[ \left(\frac{P_{\rm b}}{2 \pi} \right)^{2} \, G \left( M_{\rm ns} + M_{\rm com}\right) \right]^{1/3}$. Thus, to estimate $\mathcal R$, one needs to know various parameters like $a_{\rm ns}$ (or $P_{\rm b}$), $e$, $f$, $M_{\rm ns}$, and $M_{\rm com}$. The standard value of $M_{\rm ns}$ is $1.4 \, {\rm M_{\odot}}$.

Moreover, the path length of the neutron star within the asteroid belt can be estimated if the location of the belt in the orbit can be determined. This path length is the arc-length of the orbit inside the belt:

\begin{equation}
s = \int_{f_1}^{f_2} \sqrt{r_{\rm ns}^2 + \left( \frac{dr_{\rm ns}}{df}\right)^2} \, df ,
\end{equation} where $f_1$ and $f_2$ are true anomalies of the neutron star when it enters and exits the belt, and $r_{\rm ns}$ is the magnitude of the radius vector of the neutron star in its orbit, defined as:

\begin{equation}
r_{\rm ns} (f) = \frac{a_{\rm ns} (1-e^2)}{(1+e \cos f)}.
\end{equation}

In the next section, I fit  reported detections and non-detections of bursts from the direction of FRB121102 to extract a value of $P_{\rm b}$, obtain realistic values of $\mathcal R$ for a wide range of other parameters, and estimate values of the path-length.

\section{Application of the model for the case of FRB121102} \label{sec:applicationmodel}

Epochs of detections and non-detections of bursts from the direction of FRB121102 are noticeable in the compilation of the published results in Table \ref{tab:frbstat}. In the present model, the epochs of non-detections (09-December-2015 to 01-February-2016 as mentioned in \citet{Scsh16} and 01-May-2016 to 27-May-2016 as mentioned in \citet{clw17}) can be explained as the neutron star being in the orbital phases out of the asteroid belt, while the epochs of detections are the times when the neutron star was inside the belt. The neutron star would pass through one of the turning points of the orbit between two such epochs of detections, i.e. it would exit the belt, pass through a turning point and enter the belt again (at another location).

Table \ref{tab:frbstat} shows that the mid point of the last epoch of detection was at a separation of 295 days from that of the previous epoch of detection, which was again separated from the epoch of detection preceding it by 175 days. This fact suggests that the orbital period of the neutron star is around 470 days (295+175 days), and it takes around 295 days for the neutron star to travel from one mid-location in the belt to another through a turning point and around 175 days to return to the first location through the other turning point.

By changing the above intervals slightly (1 day each)\footnote{Mid points of epochs of detections can be slightly different from the times the neutron star comes to the middle of the belt as it is not necessary that the asteroids would start plunging as soon as the neutron star enters the belt.}, I can fit all the epochs shown in Table \ref{tab:frbstat}. I call the mid point of first epoch of detection as position-1 which is MJD 56233 and can be considered a location well inside the belt. I get subsequent times for the neutron star to be inside the belt as: position-a = position-1$+$174 days = MJD 56407, position-b = position-a$+$294 days = MJD 56701, position-c = position-b$+$174 days = MJD 56875, position-2 = position-c$+$294 days = MJD 57169, position-3 = position-2$+$174 days = MJD 57343, position-4 = position-3$+$294 days = MJD 57637. These are the positions when the neutron star was inside the belt experiencing asteroid infalls that triggered series of FRBs. Note that, these positions are very close to the mid-points of each epoch. This proximity supports the validity of the present model. None of these locations falls in the epochs of reported non-detections. This fit also allows me to assume $P_{\rm b} = 468$ days.

Following the same logic, the next two epochs when the neutron star will pass the asteroid belt are position-x1 = position-4+174 days = MJD 57811 (27-February-2017) and position-x2 = position-x1 + 274 = MJD 58105 (18-December-2017). Bursts near those times are expected.

Presently there is no reported data for position-a, position-b, and position-c although the present model expects bursts during those epochs. It would be interesting if the observing team clarifies whether any sensitive radio telescope was pointed to this direction during those epochs. However, observed data lacking any noticeable burst would not rule out the present model. If the observed data were contaminated by RFI then it would be difficult to identify bursts. The bursts can be even of very low luminosity. The luminosity of bursts in this model is given as $L \simeq 2.63 \times 10^{40} \, m_{18}^{8/9} $ if standard values for other parameters are used (Table 2 of \citet{dwwh16}) where $m_{18}$ is the mass of the asteroids in the unit of $10^{18}$ gm. The luminosity would decrease by a factor of 7.7 if the mass of the asteroid is smaller by one order of magnitude and by a factor of 59.9 if the mass of the asteroid is smaller by two orders of magnitude. So it is possible to have a series of very faint bursts if all of the asteroids that fell onto the neutron star during a particular passage are of low mass. The last possibility is the true absence of bursts. This can be the case if the asteroid belt whose constituents are also orbiting around the companion of the neutron star, has some local voids and the neutron star crossed the belt through those voids.  At present, it is not possible to favor one scenario over the others - more published data motivated to test these possibilities are essential.

In the next section, I estimate $\mathcal{R}$ and $s$ for different test binaries by varying $e$ and keeping $P_{\rm b}$ fixed at the value of $468$ days. Neutron stars can exist in such wide binaries - two binary pulsars with similar values of $P_{\rm b}$ are PSR B1800-27 with $P_{\rm b} = 407$ days, $e=0.0005$, $M_{\rm c,~med} = 0.17~{\rm M_{\odot}}$ \citep{jml95} and PSR J0214+5222 with $P_{\rm b} = 512$ days, $e=0.005$, $M_{\rm c,~med} = 0.48~{\rm M_{\odot}}$ \citep{slr14}. Note that, for such wide period binaries, the general relativistic effects (over a few orbits as discussed here) can be ignored.

\begin{deluxetable*}{lllrr}
\tablecaption{All bursts detected so far from the direction of FRB121102.  \label{tab:frbstat}}
\tablecolumns{6}
\tablenum{1}
\tablewidth{0pt}
\tablehead{
\colhead{Calendar-date} & 
\colhead{MJD} &
\colhead{no. of bursts} &
\colhead{ref}  & \colhead{interpretation} \\
\colhead{} & \colhead{} & 
\colhead{} & \colhead{} & \colhead{}
}
\startdata
2012-11-02   & 56233 &  1 & \citet{Spsh16} & inside the belt  \\
 & (position-1 is 56233)     &  &  &    \\
  &  &   &  &      \\
2012-12-09   & 56270   &  0 & \citet{Spsh16}  & out of the belt  \\
2015-05-09   &   &   0 & \citet{Spsh16}  & out of the belt \\
 &  &   &  &      \\
2015-05-17 & 57159 &   2 & \citet{Spsh16} &  inside the belt \\
2015-06-02 &  	57175 &   8 & \citet{Spsh16} & still inside the belt  \\
\multicolumn{3}{l}{(mid-point is MJD 57167; position-2 is 57169)} &    &   \\
 &  &   &    &    \\
2015-11-13 to 2015-11-19 &	57339 to 57345  &   5 & \citet{Scsh16}  & again inside the belt \\
\multicolumn{3}{l}{(mid-point is MJD 57342; position-3 is 57343)} &   &    \\
   &  &   &    &    \\
2015-12-09 to 2016-02-01 & 57365 to 57419 &   0 & \citet{Scsh16}  & out of the belt  \\
2016-04-27 to 2016-07-27 &  &   0 & \citet{clw17}  & out of the belt \\
  &  &   &  &      \\
2016-08-23 to 2016-09-20 & 57623 to  57651 &   13 (9$+$4) & \citet{clw17}  & inside the belt  \\
\multicolumn{3}{l}{(mid-point is MJD 57637; position-4 is 57637)}    & \citet{mph17} &    \\
\enddata
\end{deluxetable*}

\section{Demonstration with FRB121102} \label{sec:res}

Values of $f$ at different positions are needed to calculate $v_{\rm b, ns}$ (hence $\mathcal{R}$) and $s$. one can estimate $f$ solving Kepler's equations if in addition to $P_b$ and $e$, the time of the periastron passage is also known. I proceed with logical guesses for the periastron passage time as discussed below.

As the interval (174 days) between position-1$-$position-a, position-b$-$position-c, and position-2$-$position-3 is shorter than that between position-a$-$position-b, position-c$-$position-2, and position-3$-$position-4 (294 days), it is natural to think that the neutron star passed through the periastron during the interval for the first group. But the reverse is also possible if the asteroid belt cuts the orbit very close to the apastron. I call the first scenario `case-A' and the second scenario `case-B'. In the next two subsections, I explore `case-A' and `case-B' successively.

\subsection{Case-A}

I choose an arbitrary time, MJD 56320 between position-1 and position-a as the time of the periastron passage. I solve Kepler's equations for two very different values of $e$, one moderately high (0.5) and the other sufficiently low (0.001). I find that for $e=0.5$, position-1, position-b, position-2, and position-4 are at $f=235.4^{\circ}$, and position-a, position-c, and position-3 are at $f=124.7^{\circ}$. For $e=0.001$, position-1, position-b, position-2, and position-4 are at $f=292.97^{\circ}$, and position-a, position-c, and position-3 are at $f=67.10^{\circ}$. 

It is obvious that the values of $f$ at different positions would change if the time of the periastron passage was different, e.g., if the periastron passage was on MJD 56300 (still between position-1 and position-a); position-1, position-b, position-2, and position-4 would have $f=249.3^{\circ}$, and position-a, position-c, and position-3 would have $f=135.5^{\circ}$ for $e=0.5$. For $e=0.001$, position-1, position-b, position-2, and position-4 would have $f=308.4^{\circ}$, and position-a, position-c, and position-3 would have $f=82.5^{\circ}$.

%%%%%%%%

\subsection{Case-B}

Now I assume that one periastron passage of the neutron star was between position-a and position-b. For the purpose of demonstration, I choose MJD 56554 as the time of the periastron passage.

For $e=0.5$, position-1, position-b, position-2, and position-4 are at $f=152.31^{\circ}$, and position-a, position-c, and position-3 are at $f=207.73^{\circ}$. For $e=0.001$, position-1, position-b, position-2, and position-4 are at $f=113.18^{\circ}$, and position-a, position-c, and position-3 are at $f=246.89^{\circ}$.

%%%%%%%%%%%%%%

\subsection{Case-A and Case-B together}

Using Eqn. \ref{eq:vns}, I calculate $v_{\rm ns}(f_i)$ for each value of $f=f_i$ (in intervals of $0.1^{\circ}$) when the neutron star is inside the belt, and then compute the weighted average as $v_{\rm ns, avg} = \sum_i w_i v_{\rm ns}(f_i) / \sum_i w_i $ where $w_i = \left( 1 - e \right)^2 /\left( 1 +e \cos f_i \right)^2 $ is a weight factor corresponding to the relative duration the neutron star spends at a particular value of $f_i$ \citep{blw13}. This $v_{\rm ns, avg}$ is used while calculating $\mathcal R$ for different cases. I use the standard value for the number density of asteroids in the belt, i.e. $N_{a} / (\eta_1 \eta_2 R_{\rm belt}^3 ) = 1.5625 \times 10^{11} ~{\rm AU^{-3}}$  \citep{dwwh16}. Resulting values of $\mathcal R$ are consistent with the observed rate $\sim 3 ~{\rm h^{-1}}$ for the wide range of parameters I chose ($e$, $M_{\rm com}$, and the time of the periastron passage). This fact again supports the validity of the present model.

Table  \ref{tab:frbstat} shows that the longest burst period was around position-4, during MJD 57623 to MJD 57651 (28 days), so the neutron star was inside the belt for at least this time-span. I estimate arc-lengths around position-4 for different choices of $M_{\rm com}$, both for case-A and case-B. Table \ref{tab:arclengths1} shows that the arc-length, i.e., the minimum extent of the belt along the path of the neutron star varies between 6.7 to 37 million kilometers (0.04$-$0.25 AU). 

\begin{figure*}[h!]
\figurenum{1}
\plottwo{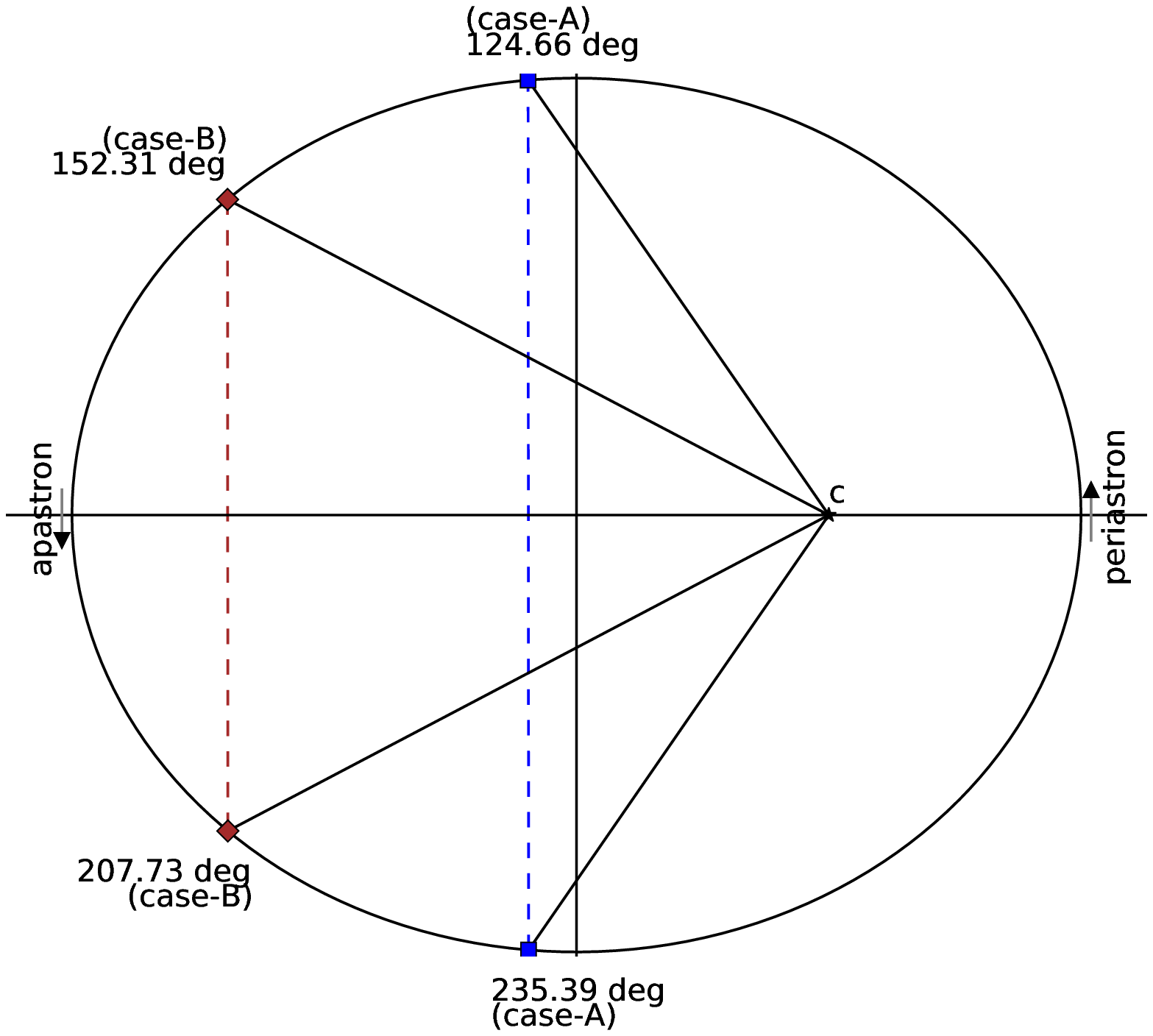}{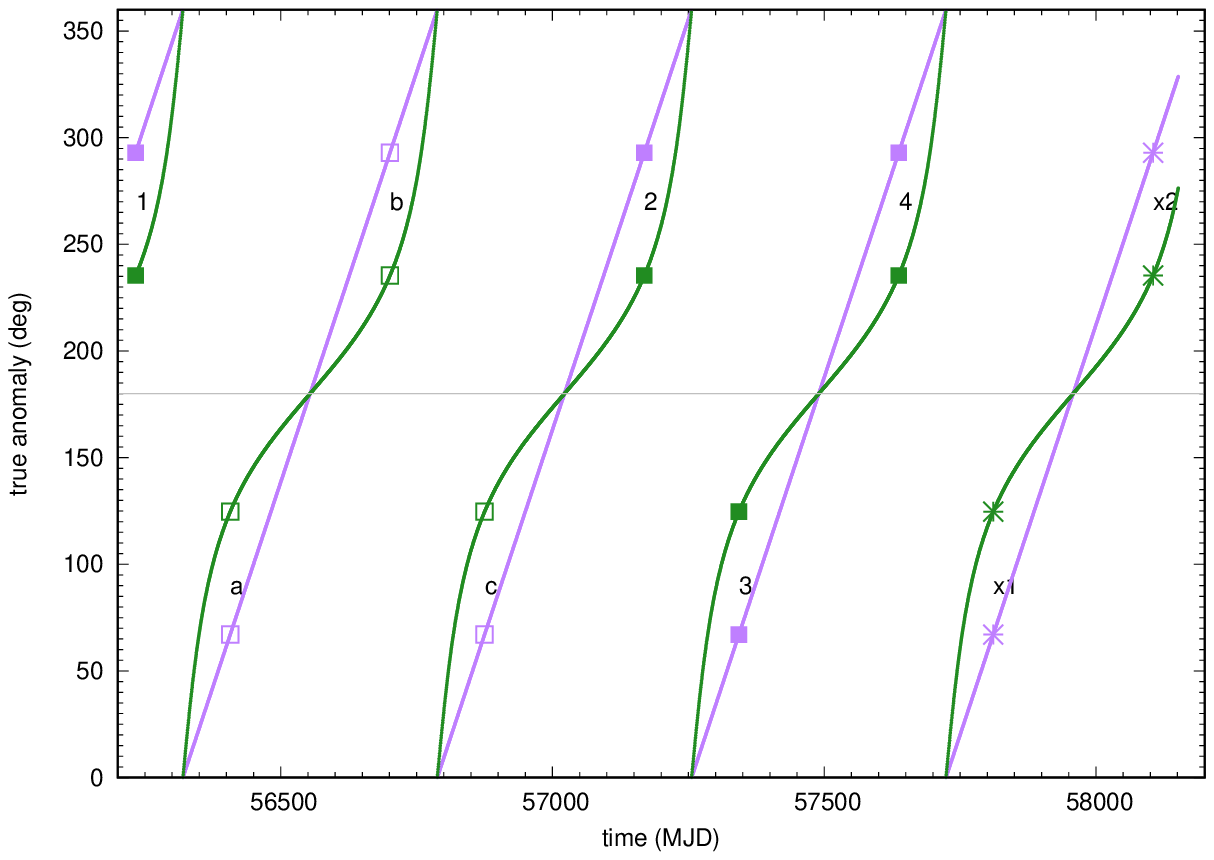}
\caption{Left-panel: Locations on the orbit (of $e=0.5$) where the neutron star crosses the asteroid belt for the periastron passage on MJD 56320 (blue squares - case-A) and on MJD 56554 (brown diamonds - case-B) with values of the true anomaly shown. The companion at one of the foci is marked as `c'. Right-panel: True anomaly (in degree) versus time (in MJD) for a neutron star binary of $P_{\rm b} = 468$ days, and the time of the periastron passage as MJD 56320 (case-A). The green lines (curved) are for $e=0.5$, while the purple lines (almost straight) are for $e=0.001$. Filled squares denote the locations when bursts were detected. Unfilled squares are the expected epochs of bursts in the past (position-a, position-b, and position-c; see section \ref{sec:model}). The asterisks (``$\ast$") are the next two epochs (MJD 57811 and MJD 58105) when the neutron star will be inside the belt.  \label{fig:bin1}}
\end{figure*}

Now I demonstrate the orbital geometry and location of the neutron star in the orbit for sample cases in Fig. \ref{fig:bin1}. In the left panel, locations on the orbit where the neutron star crosses the asteroid belt for the periastron passage on MJD 56320 (blue squares, case-A) and on MJD 56554 (brown diamonds, case-B) are shown for $e=0.5$. The right panel shows the variation in $f$ with time for case-A. The green lines (curved) are for $e=0.5$, while the purple lines (almost straight) are for $e=0.001$. Filled squares denote the epochs (and locations in the orbit) when bursts were detected. Unfilled squares are the expected epochs of bursts in the past (position-a, position-b, and position-c; see section \ref{sec:model}). The asterisks (``$\ast$") are the next two epochs (MJD 57811 and MJD 58105) when the neutron star will be inside the belt.

%%%%%%%%%%%%%%%%%%%%%%

\begin{deluxetable*}{llccccccc}
\tablecaption{Calculation of arc-lengths of the orbit around location-4, average orbital speed of the neutron star in those arcs, and the rate of asteroid plunges for some sample cases. Canonical values for the mass and the radius of the neutron star have been used, i.e. $M_{\rm ns} = 1.4 ~ {\rm M_{\odot}}$ and $R_{\rm ns} = 10$ km. \label{tab:arclengths1}}
\tablecolumns{9}
\tablenum{2}
\tablewidth{0pt}
\tablehead{
\colhead{periastron passage} & 
\colhead{$e$} &
\colhead{true anomalies } &
\colhead{$M_{\rm com}$} &
\colhead{$a_{\rm R}$}  & \colhead{$a_{\rm ns}$} & 
\colhead{s} & \colhead{$v_{\rm b, ns, avg}$} &
\colhead{$\mathcal R$}  \\
\colhead{} & 
\colhead{} &
\colhead{at MJD 57623 and MJD 57651} &
\colhead{} & \colhead{}  &
\colhead{}  & \colhead{} & 
\colhead{} \\
\colhead{(MJD)} & 
\colhead{} &
\colhead{($f_1$, $f_2$)} &
\colhead{(${\rm M_{\odot}}$)} &
\colhead{(km)}  & \colhead{(km)} &
\colhead{(km)} & \colhead{(${ \rm km ~ s^{-1}}$)} &
\colhead{(${\rm h^{-1}}$)}  
}
\startdata
  &      &   & 0.2 & $2.1 \times 10^{8}$ &  $ 2.6 \times 10^{7}$  & $ 9.2 \times 10^{6}$ & 86.7 & 3.2  \\
  & 0.5   & ($227.6^{\circ}$, $244.7^{\circ}$)  & 0.6 & $2.2 \times 10^{8}$  &  $6.7 \times 10^{7}$  & $2.4 \times 10^{7}$ & 60.3 & 4.5  \\
 &   &   & 1.0 & $2.4 \times 10^{8}$ &  $9.8 \times 10^{7}$  & $3.5 \times 10^{7}$ & 54.4 & 5.0 \\ 
56320 (case-A) &   &    &   &  &  &    &  &  \\
  &   &   & 0.2 & $2.1 \times 10^{8}$ &  $2.6 \times 10^{7}$  & $9.7 \times 10^{6}$ & 90.8 & 3.0 \\
& 0.001 &  ($282.2^{\circ}$, $303.8^{\circ}$) & 0.6 & $2.2 \times 10^{8}$ &  $6.7 \times 10^{7}$  & $2.5 \times 10^{7}$ & 63.1 & 4.3  \\      
 &   &   & 1.0 &  $2.4 \times 10^{8}$ & $9.8 \times 10^{7}$   & $3.7 \times 10^{7}$ & 56.9  & 4.8  \\    \hline  
 &   &   & 0.2 & $2.1 \times 10^{8}$ &  $2.6 \times 10^{7}$  & $6.7 \times 10^{6}$ & 63.3  & 4.3  \\
 & 0.5  & ($147.0^{\circ}$, $157.3^{\circ}$)  & 0.6 & $2.2 \times 10^{8}$ &  $6.7 \times 10^{7}$  & $1.7 \times 10^{7}$ & 44.0 & 6.2 \\
 &      &   & 1.0 &  $ 2.4\times 10^{8}$ & $9.8 \times 10^{7}$   & $2.6 \times 10^{7}$ & 39.7 & 6.9  \\ 
56554 (case-B) &   &   &   &  &  &    &  &  \\
   &       &   & 0.2 & $2.1 \times 10^{8}$ &  $2.6 \times 10^{7}$  & $9.7 \times 10^{6}$ & 90.68 & 3.0  \\
  & 0.001 & ($102.4^{\circ}$, $123.9^{\circ}$)  & 0.6 & $2.2 \times 10^{8}$ &  $ 6.7 \times 10^{7}$  & $2.5 \times 10^{7}$ & 63.1 & 4.3  \\      
  &       &   & 1.0 &  $2.4 \times 10^{8}$ &  $9.8 \times 10^{7}$  & $3.7 \times 10^{7}$ & 56.9 & 4.8  \\     
\enddata
\end{deluxetable*}

%%%%%%%%%%%%

\section{Discussions} \label{sec:discuss}

Detection of bursts close to the predicted epochs, i.e. around 27-February-2017 and 18-December-2017 would be a stronger support of the present model. Ruling out via non-detection would better be done only after a number of successive failures, as I have already argued for the absence of bursts.

A future discovery of an FRB with only one active epoch of several bursts can be explained either by a very wide binary in which the neutron star has crossed the asteroid belt around its companion only once (after such burst searches have been initiated) or by the passage of an isolated neutron star through an asteroid belt around another star.

Non-repeating FRBs can occur under different circumstances, e.g., (i) the asteroid belt is around the neutron star itself and a chance fall of an asteroid from that belt onto the neutron star leads to a single burst, (ii) a neutron star grazing an asteroid belt around another star, or (iii) a binary neutron star grazing an asteroid belt around its companion (possible if the orbit of the neutron star and the asteroid belt are non-coplanar) and the orbit is so wide that only one such grazing event has occurred so far. Scenario (i) seems to be the most likely as the scenarios (ii) and (iii) are very restrictive - the neutron star should not pass through the middle of the belt, it must graze the belt so that it would suffer only one impact during each grazing. Under scenario (i), the rate ($10^3 - 10^4 ~ {\rm sky^{-1} \, day^{-1}}$) calculated by \citet{gh15} holds valid. However, as the other two scenarios cannot be ruled out, we note that the first FRB (Lorimer burst) occurred on MJD 52146 which was almost 10 years ago (5631 days ago if I choose the current date as 2017-01-24); and it is possible for a neutron star to have an orbital period larger than 5631 days - as an example, the orbital period of PSR J2032+4127 is 8578 days and the eccentricity is 0.93 \citep{lsk15}.

Thus, a diverse variety of FRBs can be explained with this model without changing the basic physics behind generation of bursts, only by considering different configuration of the asteroid belt.

One of the non-repetitive FRBs, FRB 131104 has been recently associated with a bright gamma-ray transient \citep{dfm16}. Although soft-gamma ray emission is possible after the asteroid$-$neutron star impact \citep{dwwh16}, the gamma-ray flux $F_{\gamma} =  \dot{E}_{\rm G} / 4 \pi d_L^2$ for $d_L = 3.5$ Gpc\footnote{http://www.astronomy.swin.edu.au/pulsar/frbcat/} is too low ($ \sim 10^{-16}~{\rm erg~ s^{-1} cm^{-2}}$) if the values of the parameters for the neutron star and the asteroids are as usual \citep[see equation 3 for the expression of $ \dot{E}_{\rm G}$.]{dwwh16}. However, because of the uncertainties in both of the claimed association and the estimation of $d_L$ (mainly due to the uncertainties in the models of the dispersion measure for both of the interstellar medium and the intergalactic medium), it is not yet possible to exclude the present model being the cause of this FRB.

This model will remain valid even in the case of a future detection of a low dispersion measure FRB, i.e. an FRB in the Galaxy, as a binary neutron star with an asteroid belt around the companion or a neutron star having an asteroid belt around itself can very well exist in the Galaxy.

\acknowledgments

The author thanks the anonymous reviewers for useful comments which improved the manuscript.

\end{document}